\documentclass{nature}
\usepackage{graphicx}
\usepackage{epsfig}

\begin{document}

\title{Thermal Conductivity Minimum: A New Water Anomaly}

\maketitle

\bigskip

\noindent {\bf Pradeep Kumar and H. Eugene Stanley}

\noindent {\it Center for Polymer Studies and Department of Physics,

\noindent Boston University, Boston, MA 02215 USA}

\bigskip

\bigskip

\noindent Many anomalies of water have been discovered. Some concern the
static thermodynamic properties of water, such as the increase of
compressibility and specific heat upon decreasing the
temperature. Others concern the dynamic properties, such as the
breakdown of the Stokes-Einstein relation in supercooled water
\cite{chenmallamace2006} and the non-Arrhenius to Arrhenius dynamic
crossover \cite{Liu04,xuPNAS}.

The thermal conductivity $\Lambda$ is the proportionality constant
between heat flux and temperature gradient that causes the flux. Upon
decreasing temperature below its maximum at $400$~K
\cite{thermExp2,thermExp1}, $\Lambda$ decreases. However no data have
been reported below $\approx 268$~K. Using molecular dynamics computer
simulations of water down to $T=220$~K, we find the surprising result
that the $\Lambda$ displays displays a minimum at lower temperatures,
$T_{\rm min}=250$~K. Our simulations are for the TIP5P model of water
which is known to reproduce experimental data well at ambient
pressures \cite{YamadaXX,Paschek05}.

We first perform equilibrium simulations at atmospheric pressure for
temperatures from $300$~K down to $220$~K in the $NPT$ ensemble using
the Berendsen thermostat and barostat with a time step of $1$~fs. We
then run the equilibrated configurations at different $T$ in the $NVE$
and $NVT$ ensembles to generate the results from which we calculate
$\Lambda$. To calculate $\Lambda$, we use two different methods: (i)
the Green-Kubo relation between $\Lambda$ and the energy current
correlation function \cite{bookAllen}, and (ii) the M\"uller-Plathe
algorithm \cite{muller}.  Our values for $\Lambda$ obtained from both
methods differ from each other by less than $15$\%.

Figure~1(a) shows that $\Lambda$ decreases with decreasing $T$,
reaching a minimum at $T_{\rm min}\approx 250$~K, and increases upon
further decrease in $T$.  A clue to the possible interpretation of
this surprising result is provided by the observation that the
temperature at which $\Lambda$ displays a {\it minimum\/} is the same
temperature at which the specific heat $C_P$ displays a {\it
maximum\/} [see Fig.~1(b)].

In fact, a maximum in $C_P$ occurs upon crossing the locus of maximum
correlation length, the Widom line $T_W(P)$ \cite{xuPNAS}, emanating
from the hypothesized liquid-liquid critical point
\cite{Poole1}. Below $T_W(P)$, water becomes locally more tetrahedral
\cite{kumarPRL}, and the local structure of liquid water below
$T_W(P)$ more resembles ice $I_h$. Hence $\Lambda$ should increase
with further decrease in temperature below $T_W(P)$.

Finally we calculate the thermal diffusivity $D_T$, related to $\rho$,
$\Lambda$ and $C_P$ as
\begin{equation}
D_T = \frac{\Lambda}{\rho C_P}.
\end{equation}
Figure~1(c) show that $D_T$ decreases very slowly upon decreasing $T$
and rises sharply upon further decreasing $T$ below the temperature of
maximum $C_P$.

In summary, we report a new anomaly of water, a minimum in the thermal
conductivity. Our findings are of special interest since they support
the presence of a liquid-liquid phase transition in water
\cite{Poole1}.

\begin{figure}
\begin{center}
\includegraphics[width=8cm,height=16cm]{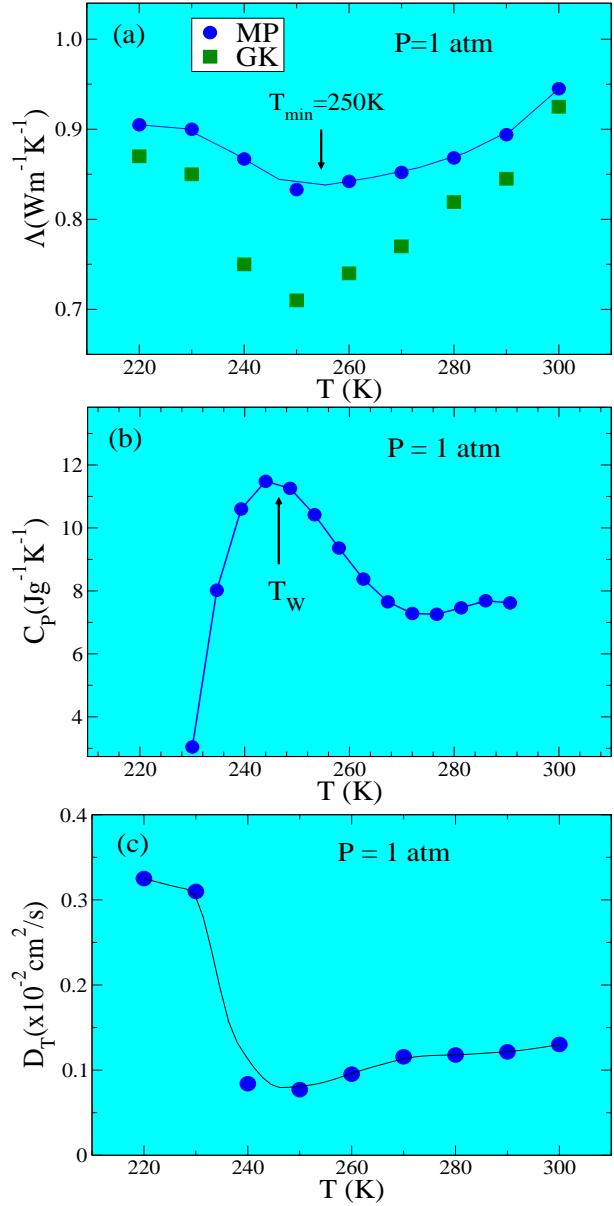}
\caption{(a) Temperature dependence of the thermal conductivity
  $\Lambda$ of TIP5P water, showing a minimum at $T_{\rm min} \approx
  250$~K. Values of $\Lambda$ obtained from the M\"uller-Plathe
  algorithm are shown as filled circles and $\Lambda$ obtained via the
  Green-Kubo relation are shown as filled squares. (b) Temperature
  dependence of the specific heat $C_P$ showing a maximum at
  $T_W\approx250$~K. (c) Temperature dependence of the thermal
  diffusivity $D_T=\Lambda/\rho C_P$, showing a sharp increase below
  $T_W(P)$.}
\end{center}
\label{fig:fig1}
\end{figure}


\begin{addendum}
 \item We thank S.-H. Chen for helpful comments and the NSF Chemistry
 Program for support.
 \item[Competing Interests] The authors declare that they have no
competing financial interests.
 \item[Correspondence] Correspondence and requests for materials
should be addressed to Pradeep Kumar (email:
pradeep@buphy.bu.edu).
\end{addendum}


\begin{thebibliography}{1}

\bibitem{chenmallamace2006} Chen, S.-H. et al.
{\it Proc. Natl. Acad. Sci. USA} {\bf 103}, 12974--12978 (2006).


\bibitem{Liu04} Liu, L., Chen, S.-H., Faraone, A., Yen, C.-W \& Mou, C.-Y.
{\it Phys. Rev. Lett.} {\bf 95}, 117802 (2005).

\bibitem{xuPNAS} L. Xu et al.
{\it Proc. Nat. Acad. Sci. USA} {\bf 102}, 16558 (2005).


\bibitem{thermExp2} IAPWS Release on the {\it Thermal Conductivity of
Ordinary Water Substance}, IAPWS Secretariat (1998).



\bibitem{thermExp1} E. H. Abramson, E. H., Brown, J. M. \& Slutsky,
L. J.  {\it J. Chem. Phys.} {\bf 115}, 10461--10463 (2001).

\bibitem{YamadaXX} Yamada, M.,  Mossa, S., Stanley, H. E. \& Sciortino, F.
{\it Phys.~Rev.~Lett.} {\bf 88}, 195701 (2002).

\bibitem{Paschek05} Paschek, D. {\it Phys. Rev. Lett.} {\bf 94}, 217802
  (2005).

\bibitem{bookAllen} Allen, M. P. \& Tildesley, D. J. {\it Computer
Simulations of Liquids\/} (Oxford Science Publications, Oxford, 1989).

\bibitem{muller} M\"uller-Plathe, F. {\it J. Chem. Phys.} {\bf 106},
6082--6085 (1997).

\bibitem{Poole1}Poole, P. H., Sciortino, F., Essmann, U. \& Stanley,
H. E. {\it Nature\/} {\bf 360}, 324--328 (1992).

\bibitem{kumarPRL} Kumar, P. et al.,
{\it Phys. Rev. Lett.} {\bf 97}, 177802 (2006).

\end{thebibliography}
\end{document}